\documentclass[aps,pra,reprint,groupedaddress,showpacs]{revtex4-1}
\usepackage{amsmath,amsfonts,natbib,verbatim}

\usepackage[pdftex]{graphicx}
\usepackage{epstopdf}
\usepackage{color}

\newcommand{\ket}[1]{|#1\rangle}
\newcommand{\bra}[1]{\langle#1|}
\newcommand{\avg}[1]{\langle#1\rangle}

\begin{document}

\title{Sub-Doppler Cavity Cooling Beyond The Lamb-Dicke Regime}

\author{Boon Leng Chuah}
\email{cqtcbl@nus.edu.sg}
\affiliation{Centre for Quantum Technologies and Department of Physics, National University of Singapore, 3 Science Drive 2, 117543 Singapore}
\author{Nicholas C. Lewty}
\affiliation{Centre for Quantum Technologies and Department of Physics, National University of Singapore, 3 Science Drive 2, 117543 Singapore}
\author{Radu Cazan}
\affiliation{Centre for Quantum Technologies and Department of Physics, National University of Singapore, 3 Science Drive 2, 117543 Singapore}
\author{Murray D. Barrett}
\affiliation{Centre for Quantum Technologies and Department of Physics, National University of Singapore, 3 Science Drive 2, 117543 Singapore}

\date{\today}

\begin{abstract}

We investigate the dynamics of cavity cooling of a single ion beyond the Lamb-Dicke regime and demonstrate a cooling limit of approximately 50\% of the Doppler temperature using a high finesse cavity for the first time. We also study the statistical properties of the ion-cavity emission and present a model that maps the phonon states to the photon states. With this model, we explain the super-Poissonian photon distribution observed in our experiments and propose a method to estimate the ion temperature in a real-time manner based on the statistical behavior of the photon emission from the cavity.

\end{abstract}


\maketitle
\section{Introduction}

Trapped ions have proven to be a promising system for quantum information processing (QIP) applications \cite{wineland,home,wineland2} with all of the experimental requirements having been demonstrated. In recent years experimental efforts have been made towards establishing entanglement between distant traps \cite{blinov2004observation, moehring2007entanglement, olmschenk2009quantum}.  This remote entanglement serves as a building block for quantum networks and provides a path to large scale QIP. While ions are good candidates for stationary processes due to their long lived internal states (e.g. quantum memory), photons are the best carriers of quantum information between physically separated sites \cite{cirac,duan2,gheri}. Thus, an ion-photon interface is important for the development of large scale QIP. An ideal system for such an interface is based on an ion trapped within a high finesse cavity \cite{mundt,keller,stute}. The cavity enhances the interaction between the ion and a single photon, and enables efficient extraction of emitted photons. Proposed applications of trapped ion-cavity systems in QIP include quantum repeaters \cite{briegel,sangouard}, entanglement of distant ions \cite{browne,simon,ye} and quantum logic gates \cite{semiao,zheng,zheng2,zheng3}. To date, remarkable advancements have been made: single photon sources \cite{keller,barros}, single ion lasers \cite{dubin} and ion-photon entanglement \cite{stute} have all been demonstrated with trapped ion-cavity setups.

In addition to QIP applications, a cavity also provides other useful functions for a trapped ion system such as enhanced photon collection efficiency \cite{Sterk}, detection and minimization of excess micromotion \cite{chuah2} and a means for cooling ions \cite{chuang}. In particular, a notable property of cavity cooling is that it can, in principle, be performed without affecting the logical information stored in the atomic internal states \cite{vuletic2,vuletic}. In addition, cavity cooling techniques can be effective well below the Doppler limit.  This is important for heavier ions for which Doppler cooling typically results in a large mean vibrational state, making it difficult for further sideband cooling.  Cavity cooling of neutral atoms has been demonstrated and significant experimental milestones such as ground state cooling have been achieved \cite{maunz,boozer}. However, for ions, cavity experiments have been limited \cite{chuang} and cooling below the Doppler limit is yet to be reported.

In this paper, we investigate cavity cooling of a single ion beyond the usual Lamb-Dicke (LD) regime in which the localization of the atom/ion is much smaller than the wavelength of the atomic transition. We demonstrate ion cooling to sub-Doppler temperatures which, to our knowledge, is the first demonstration of sub-Doppler cooling of an ion using a high finesse cavity. Motivated by the observation of super-Poissonian behavior in the ion-cavity emission, we also study the photon statistics of the intra-cavity field.  It is shown that the thermal state of the ion is reflected in photon emission from the cavity.  This potentially allows for the determination the thermal state of the ion from the statistical properties of the cavity emission.

\section{Cavity Cooling} \label{cavitycooling}

Two types of cavity cooling have been discussed in the literature, namely cavity Doppler cooling and cavity sideband cooling \cite{vuletic}. The former is used for cooling free particles (atoms/molecules) while the latter is used when they are strongly confined in the LD regime \cite{vuletic}. Here we operate in an intermediate regime where neither description is adequate.

Cavity sideband cooling of trapped atoms has been extensively studied theoretically \cite{morigi2,morigi,vuletic}. The cooling models are all derived based on the assumption that the atoms are in the LD regime, $\eta^2 \avg{n} \ll 1$, where $\eta$ is the LD parameter. In this case, the single atom cooperativity $C$ \cite{horak}, which characterizes the ratio between the scattering rate into the cavity versus that into the free-space, is unaffected by the change in the temperature of the atom. As $\eta^2 \avg{n}$ increases, the thermal dependence of the scattering rate into the cavity has to be taken into consideration. Even though our system has a maximum $\eta^2 \avg{n}$ of only $\sim0.3 $, the thermal effect is observable. Here we experimentally demonstrate that cavity cooling is well described by an adaption of the model given in \cite{morigi} to account for the dependence of the scattering rate into the cavity on the thermal state of the ion.  


\begin{figure}
\includegraphics{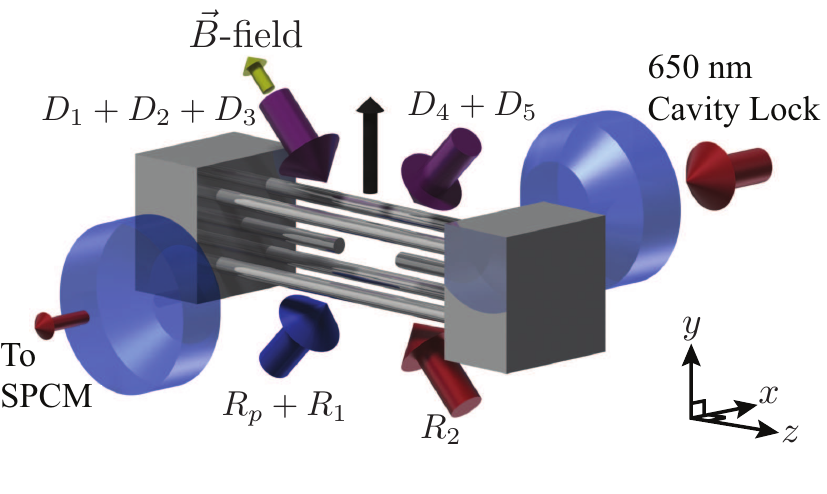}
 	 \caption{\label{iso} (Color online) The experimental setup. A single $^{138}\textrm{Ba}^{+}$ ion is trapped and coupled to a high finesse cavity. The emitted photons from the cavity are collected into a fiber-coupled single-photon counting module (SPCM). A CCD camera, interchangeable with another free-space SPCM, detects the fluorescence of the ion in the direction of the black arrow. The state manipulation and the ion cooling are performed with the $D_1-D_5$, $R_p, R_1$ and $R_2$ laser beams. See text for a detailed description of each beam. }
\end{figure}

\subsection{Setup} \label{setup}

The experimental setup, illustrated in Fig.~\ref{iso}, consists of a high finesse cavity aligned with its optical axis transverse to a linear Paul trap \cite{prestage,berkeland2,chuah}. Details of the ion trap have been reported elsewhere \cite{lewty}. Briefly, a $5.3 \, \textrm{MHz}$ RF potential with an amplitude of $125 \, \textrm{V}$ is applied via a step-up transformer to two diagonally opposing electrodes. A small dc voltage applied to the other two electrodes ensures a splitting of the transverse trapping frequencies. Axial confinement is provided by two axial electrodes separated by $2.4 \, \textrm{mm}$ and held at $33\, \textrm{V}$. Using this configuration, we achieve trapping frequencies of $2 \pi \times (1.2, 1.1, 0.40) \, \mathrm{MHz}$ for a single $^{138}\mathrm{Ba}^+$ ion.

\begin{figure}
\includegraphics{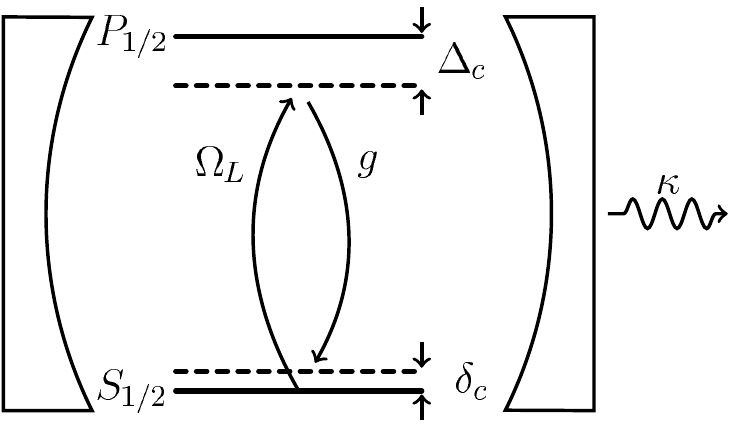}
\caption{\label{cavlevel} The transition driven by the cavity probing beam $R_p$ and the intra-cavity field. The metastable $\mathrm{D}_{3/2}$ state is omitted in the figure. $\Delta_c$ is the detuning of the laser frequency from the $\textrm{S}_{1/2} \leftrightarrow \textrm{P}_{1/2}$ transition while $\delta_c$ is the relative detuning between the laser and the cavity resonance. $\Omega_L$ is the Rabi rate of beam $R_p$. $g$ is the coupling strength between the ion and the cavity field. $\kappa$ is the cavity field decay rate. In cavity cooling experiments, the induced transition is resonant with the red sideband of the $\hat{\textbf{z}}$ vibrational state by setting $\delta_c =-\omega_z=- 400 \, \textrm{kHz}$. In photon statistics experiments, $\delta_c$ is set to zero. $\Delta_c$ is fixed at $-110\,\textrm{MHz}$ for all experiments. }
\end{figure}

Doppler cooling is achieved by driving the $6\textrm{S}_{1/2} \to 6\textrm{P}_{1/2}$ transitions at $493\, \textrm{nm}$ and repumping on the $5\textrm{D}_{3/2} \to 6\textrm{P}_{1/2}$ transitions at $650\, \textrm{nm}$ \cite{chuah,chuah2}. The $493\, \textrm{nm}$ cooling laser ($D_1$) and the $650\, \textrm{nm}$ repumping laser ($D_2$) are both red-detuned by $\approx 15\, \textrm{MHz}$ for optimum cooling. Both lasers are combined into a single optical fiber and sent into the trap along the $(\hat{\mathbf{z}}-\hat{\mathbf{y}})$ direction.  Additionally, $D_1$ and $D_2$ are linearly polarized perpendicular to a magnetic field of approximately 1.5 Gauss. This configuration avoids unwanted dark states in the cooling cycle.

The dual coated high finesse cavity is approximately $5\,\mathrm{mm}$ long and has a finesse of $\sim 8.5 \times 10^4$ at $493\,\textrm{nm}$ and $\sim 7.5 \times 10^4$ at $650\,\textrm{nm}$. The cavity is slightly birefringent with polarization modes split by $239(1) \, \textrm{kHz}$. The polarization modes happen to be aligned to within a few degrees of the $\hat{\textbf{y}}$ and $\hat{\textbf{z}}$ axis respectively. The cavity length is stabilized via the Pound-Drever-Hall technique \cite{hall} to an electro-optical modulator (EOM) sideband of a low linewidth $650 \, \textrm{nm}$ laser \cite{chuah}. Changing the EOM drive frequency allows us to tune the cavity resonance relative to the fixed frequency of the $650\,\mathrm{nm}$ locking laser. The laser frequency is referenced to a temperature stabilized zerodur cavity and is approximately $500\,\mathrm{GHz}$ detuned from the repump transition, thus having no impact on the cavity dynamics. A probe laser ($R_p$) at $493\,\mathrm{nm}$ drives the cavity induced Raman transition as depicted in Fig.~\ref{cavlevel}.  The laser is sent into the trap along the $(\hat{\mathbf{y}}+\hat{\mathbf{z}})$ direction and linearly polarized along the magnetic field direction. Additionally, the laser is red-detuned by $\approx 110\,\textrm{MHz}$ from the $\textrm{S}_{1/2} \leftrightarrow \textrm{P}_{1/2}$ transition and referenced to the fixed frequency of the $650\,\mathrm{nm}$ locking laser via a transfer cavity to ensure that it has a well defined detuning relative to the cavity resonance.

\begin{figure*}
\includegraphics[width=0.98\textwidth]{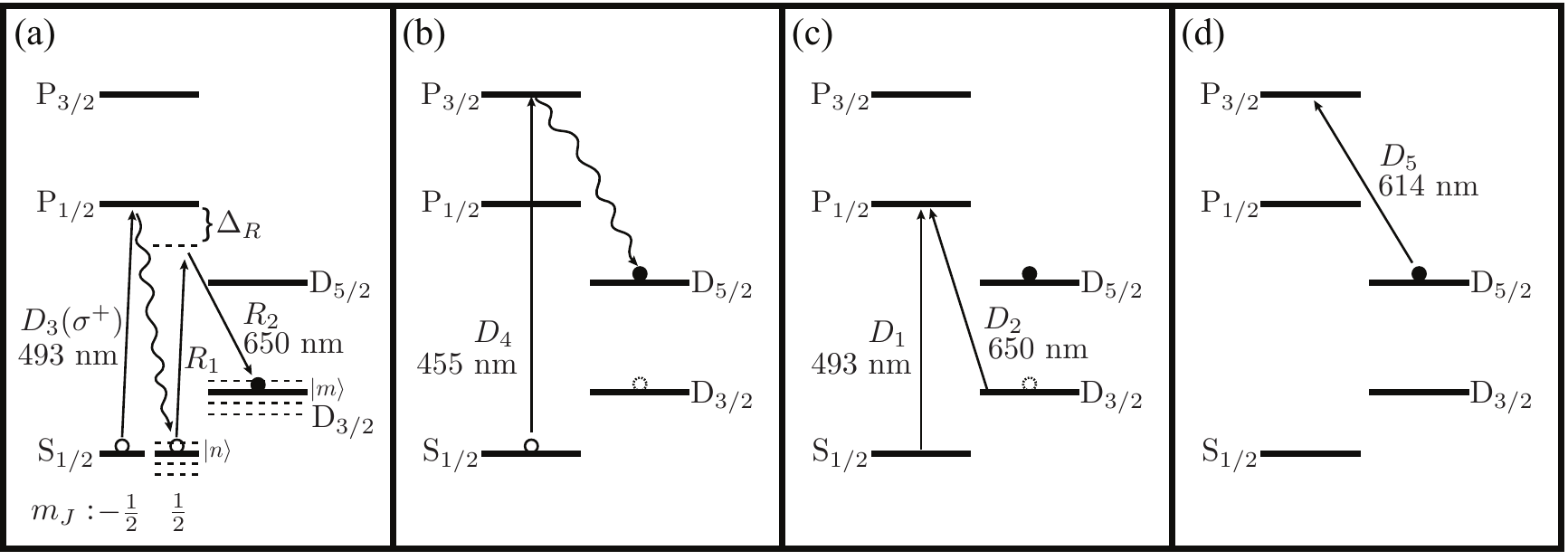}
 	 \caption{\label{temp2} The Raman transition and the state detection scheme: (a) The ion is prepared in $\textrm{S}_{1/2}\ket{m_J=1/2,n_z=n}$ using a $493\, \textrm{nm} \, \sigma^+$ laser beam, $D_3$. Afterwards, the $R_1$ and $R_2$ Raman beams drive the $\Lambda$-type Raman transition and transfer the ion to $\textrm{D}_{3/2}\ket{m_J=3/2,n_z=m}$. Both $R_1$ and $R_2$ are detuned by $\Delta_R \approx- 500 \, \textrm{GHz}$ from $\textrm{P}_{1/2}$. (b) The unsuccessfully transferred ion from $\mathrm{S}_{1/2}$ is shelved by the $D_4$ laser beam to $\mathrm{D}_{5/2}$ via $\mathrm{P}_{3/2}$ with 87\% transfer efficiency, while in 13\% of the events, the ion will end up in $\mathrm{D}_{3/2}$. (c) The Doppler cooling beams, $D_1$ and $D_2$, are turned on. If the ion is in $\mathrm{D}_{5/2}$, it is out of the cooling cycle and thus no fluorescence will be detected. (d) The repumping beam, $D_5$, depletes the population in $\mathrm{D}_{5/2}$ and moves the ion back to the Doppler cooling cycle.}
\end{figure*}

The cavity QED parameters are $(g_0,\kappa,\gamma)=2\pi \times (2.13(3), 0.172(5), 10.05(35)) \, \mathrm{MHz}$ \cite{davidson1992oscillator, stute2}, yielding a single atom cooperativity $C_0= g_0^2 / \kappa \gamma=2.64(14)$ \cite{horak}. However, the effective cooperativity is greatly reduced from this value due to the following factors: the reduced dipole matrix element for $\pi$-polarization (0.33), cavity birefringence (0.672(13) at the midpoint between the modes) and  branching ratio for $\textrm{S}_{1/2} \leftrightarrow \textrm{P}_{1/2}$ transition ($0.756(12)$) \cite{davidson1992oscillator}. The combination of these independent parameters results in an effective cooperativity of $C=0.446(21)$, valid for an ion in the LD regime.

\subsection{Ion temperature measurements} \label{tempmeasure}

A standard technique for determining the mean vibration quanta, $\avg{n}$, for a thermal state is to compare the transfer efficiencies when driving red and blue motional sidebands of a Raman transition between two internal states \cite{wineland4}. In this case the ratio of transfer efficiencies can be directly related to the mean vibrational state.  However for moderate values of $\avg{n}$, this method becomes inaccurate as the ratio saturates to unity. For larger $\avg{n}$ experimentalists have used another technique which compares Rabi flopping on the red and blue motional sidebands \cite{walther,bowler}. We have found that a more accurate and faster determination can be achieved by comparing the transfer efficiencies of several motional sidebands in the Raman spectra. As $\avg{n}$ increases higher order motional sidebands appear. Thus $\avg{n}$ in the direction of interest can be obtained from the fit of the Raman spectrum to the temperature-dependent probability of transfer from $\textrm{S}_{1/2}$ to $\textrm{D}_{3/2}$ levels.  In general this transfer probability is
\begin{eqnarray}
\label{pst}
P(\avg{n},\delta)  &=&   \sum^{\infty}_{m=-\infty}\sum^{\infty}_{k=0} \, \frac{\avg{n}^{k}}{\left( 1+\avg{n} \right)^{k+1}}  \,  \frac{\Omega_{k,k+m}^2}{\Omega_{k,k+m}^2+\delta_m^2} \,\nonumber \\
&\times& \sin^2 \Bigg( \frac{\tau_R}{2} \sqrt{\Omega_{k,k+m}^2+\delta_m^2} \, \Bigg) \, ,
\end{eqnarray}
where $\delta_m = \delta - m \, \omega$, $\tau_R$ is the Raman pulse length, $\delta$ is the detuning of the Raman frequency from the carrier transition, $\omega/2\pi$ is the trap frequency along the direction of interest and $\Omega_{i,j}$ is the effective Raman Rabi rate \cite{wineland4}
\begin{equation}
\label{ost}
\Omega_{i,j} =  \Omega_{r} \, e^{-\eta_{Rz}^2/2} \sqrt{\frac{i_{<}!}{i_{>}!}}\, \eta_{Rz}^{|i-j|} \, L^{|i-j|}_{i<}(\eta_{Rz}^2).
\end{equation}
In this equation, $\eta_{Rz}$ is the LD parameter, $L_n^\alpha(x)$ is the generalized Laguerre polynomial and the parameter $\Omega_{r}$ is given by
\begin{equation}
\Omega_{r}\,=\,\frac{\Omega_{R1} \, \Omega_{R2}}{2 \, \Delta_R}
\end{equation}
where $\Omega_{R1}$ and $\Omega_{R2}$ are the Rabi rates of the $R_1$ and $R_2$ lasers, respectively and $ \Delta_R \approx 500 \, \mathrm{GHz}$ is the detuning of the Raman lasers from the ionic resonances, $\textrm{S}_{1/2} \leftrightarrow \textrm{P}_{1/2}$ or $ \textrm{P}_{1/2} \leftrightarrow \textrm{D}_{3/2}$.

\begin{figure}
\includegraphics{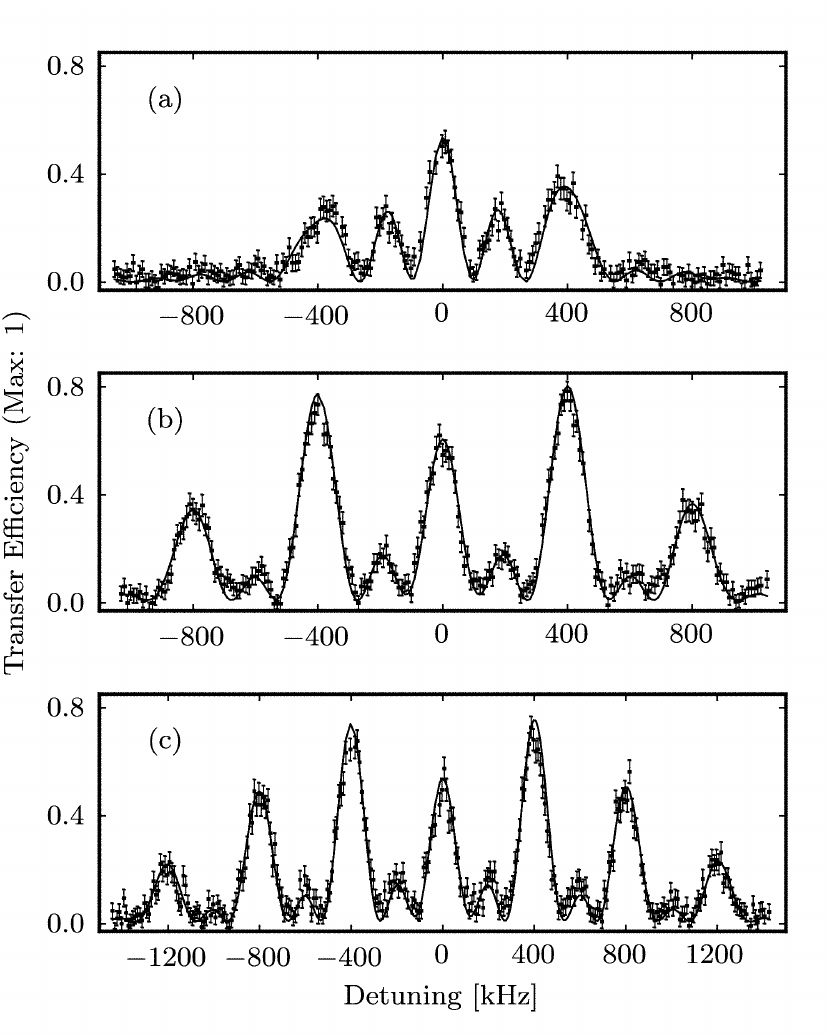}
 	 \caption{\label{temp} Three examples of Raman spectra used for inferring the temperature. The dots with error bars are the experimental data while the solid lines are the fit of Eq.~(\ref{pst}). The zero position in the horizontal axis denotes the carrier resonance. Fitted parameters: (a)  $\avg{n_z}=2.5(2)$ and  $ \Omega_r = 2 \pi \times 113(1) \textrm{kHz}$.  (b) $\avg{n_z}=30(1)$ and  $ \Omega_r = 2 \pi \times 117(2) \textrm{kHz}$. (c) $\avg{n_z}=57(2)$ and  $ \Omega_r = 2 \pi \times 118(2) \textrm{kHz}$. The reduced $\chi^2$ values of all fits are $\approx 1$.}
\end{figure}

The procedure for obtaining the spectrum is illustrated in Fig.~\ref{temp2}. Briefly, the ion is first prepared in $\textrm{S}_{1/2} \ket{m_J=1/2}$ by switching on a $493 \, \textrm{nm}$ $\sigma^+$ beam, $D_3$, and the $650 \,\textrm{nm}$ repumping beam, $D_2$, for $20 \, \mu \textrm{s}$. Afterwards, the Raman beams, $R_1$ and $R_2$, are turned on for $7 \, \mu \textrm{s}$ to transfer the ion from $\textrm{S}_{1/2}\ket{m_J=1/2}$ to $\textrm{D}_{3/2}\ket{m_J=3/2}$ via a $\Lambda$-type Raman process. The unsuccessfully transferred ion is shelved to $\mathrm{D}_{5/2}$ via $\mathrm{P}_{3/2}$ by a $455 \, \textrm{nm}$ beam, $D_4$, which is switched on for $20 \, \mu \textrm{s}$ after the Raman pulse. For state detection, the Doppler cooling beams, $D_1$ and $D_2$, are turned on for $800 \, \mu \textrm{s}$. If the ion is in $\mathrm{D}_{5/2}$, it is out of the cooling cycle and thus no fluorescence will be detected. If the ion is in $\mathrm{D}_{3/2}$, 8 photons on average will be collected into a single photon counting module (SPCM). The state detection efficiency is only $\approx 87$ \% due to the imperfect shelving process which relies on the relative branching ratio of the spontaneous decays between $\textrm{P}_{3/2} \to \textrm{D}_{3/2}$ (13\%) and $\textrm{P}_{3/2} \to \textrm{D}_{5/2}$ (87\%). After the detection, a repumping beam at $614 \, \textrm{nm}$ is turned on for $100 \, \mu \textrm{s}$ to deplete the population in $\textrm{D}_{5/2}$. To obtain a complete spectrum, the same procedure is carried out for a range of Raman frequencies which cover the relevant vibrational resonances. Examples of the obtained Raman spectra are shown in Fig.~\ref{temp}.

\subsection{Experiments} \label{cavexp}

\begin{figure}
\includegraphics{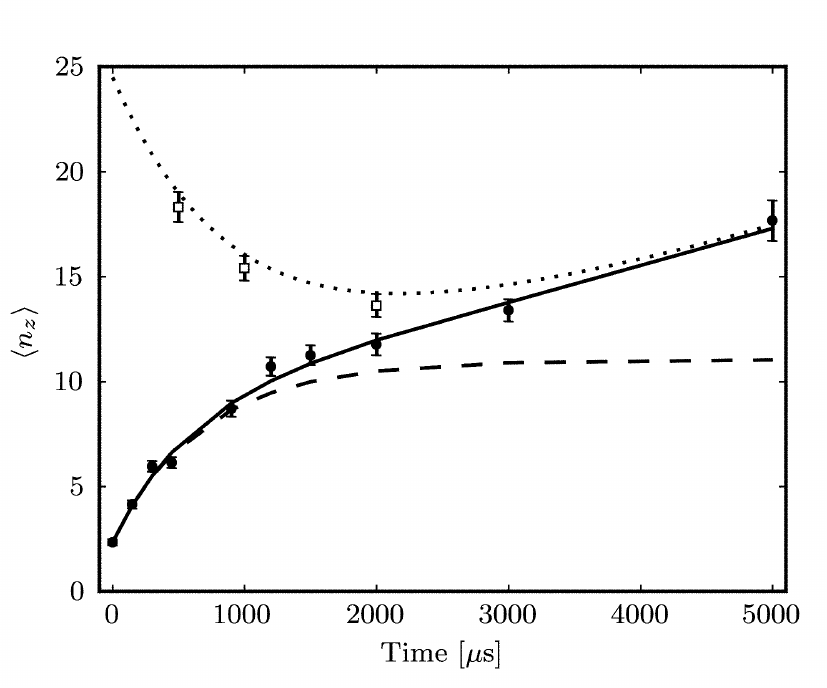}
\caption{\label{ccool} The average phonon number $\avg{n_z}$ versus cavity cooling time.  The unfilled squares are the experimental data for the cavity cooling from the Doppler limit $\avg{n_z}  \approx 25$ along the $\hat{\mathbf{z}}$ axis. The solid line is the numerical fit of Eq.~(\ref{nrate}) to the experimental data (black filled circles) obtained in the experiment where the ion phonon number is monitored at different cavity cooling duration after the initialization at $\avg{n_z}\approx2$. The values of the fitted parameters are $C=0.33(3)$, $\Omega_{L0}=2\pi \times 23(1) \textrm{MHZ}$, $\textrm{d}\avg{n_r} /\textrm{dt}=6.3(1.6)/ \textrm{ms}$ and the initial temperature $\avg{n_{z0}} = 2.4(2)$. The reduced $\chi^2$ of the fit is $\approx 1.8$. The dashed line is the simulation based on the fitted parameters from the same experiment without considering the recoil heating in the radial axes. The dotted lines is the simulation of the cavity cooling from the Doppler limit using the same parameters. The error bars on the data points are the standard deviations obtained from the temperature fits similar to those in Fig.~\ref{temp}. }
\end{figure}

In order to drive the cooling sideband along the $\hat{\mathbf{z}}$ axis, the cavity resonance is red-detuned by the axial trapping frequency $\omega_z \approx 400 \, \mathrm{kHz}$ from the ion-cavity Raman resonance. To investigate the cooling dynamics, the ion is first cooled to the Doppler temperature using a Doppler cooling pulse of $1.5\,\mathrm{ms}$.  The ion is then cavity cooled for a time $\tau$ and then the temperature is probed. The results for a range of times $\tau$ are presented in Fig.~\ref{ccool} (empty squares). A minimum temperature of $\avg{n_z}=13.6(6)$, or 54(3)\% of the Doppler limit, is achieved by applying a cooling pulse of $2\,\mathrm{ms}$.

To investigate the cooling dynamics further, another set of experiments was performed in which the ion was initially prepared with a low $\avg{n_z}$ using Raman sideband cooling \cite{monroe} immediately after the Doppler cooling stage. In our system, the sideband cooling involves a two-color Raman transition used for the ion temperature measurement, in a setup similar to \cite{chuah}. The Raman beams, $R_1 $ and $ R_2$, were arranged in such a way that the state transfer was predominantly sensitive to motion along the $\hat{\mathbf{z}}$ axis. This was achieved by having a much larger LD parameter along the $\hat{\mathbf{z}}$ axis, $\eta_{Rz}=0.15$, than that along other axes, $\eta_{Rx} \approx \eta_{Ry}=0.01$. After $3\,\mathrm{ms}$ of sideband cooling, the ion is then exposed to cavity cooling and the temperature probed as before. The resulting $\avg{n_z}$ are shown in Fig.~\ref{ccool} using black filled circles. As demonstrated, the time evolution of the ion temperature is a combination of an exponential relaxation to a steady-state temperature and a continuous, slow linear increase over time.

In order to account for our experimental results, we adapted the model given in \cite{morigi} to include for effects beyond the usual Lambe-Dicke (LD) limit. A detailed account of our model is given in Appendix~\ref{classicLD} and Appendix~\ref{recoilheating}.  Essentially, we consider the effect of the ion temperature on the scattering rate into the cavity and derive an expression for an effective single atom cooperativity as a function of $\avg{n}$ along each dimension. This effective cooperativity is substituted into the rate equations quoted from \cite{morigi}, which are subsequently used to fit the experimental data. The linear increase in $\avg{n}$ is due to recoil heating in the uncooled dimensions.  This results in a decreasing cooperativity and thus a diminished cooling rate.

In Fig.~\ref{ccool}, the solid line is the numerical fit using the rate equations Eq.~(\ref{coolandheat2}, \ref{coolandheat3}, \ref{nrate}) for a birefringent cavity. The fitted value of the cooperativity extrapolated to the LD regime, $C=0.33(3)$, is only 74(7)\% of the estimated value, 0.446(21). The discrepancy could be due to the additional heating along the ion radial direction such as recoil heating during the sideband cooling and the environmental heating caused by electronic noise.  The photon recoil contributes a heating rate of $\approx 0.05$ phonon per sideband cooling cycle. In $3\,\mathrm{ms}$ of sideband cooling process, there are 120 cooling cycles in total. Thus the phonon occupation number can increase by $\approx 6$ phonons. The environmental heating is measured to be $\approx 1$ phonon per ms. Hence, the sum of these heating effects should raise the radial vibrational occupation number by $\approx 9$ in $3\,\mathrm{ms}$ which causes an underestimation of the fitted cooperativity by $13 \%$. Including these influencing factors, the fitted value of the cooperativity extrapolated to the LD regime is $C=0.38(3)$, within $2\,\sigma$ range of the estimated cooperativity, 0.446(21).  The rest of the fitting parameters are within their respective estimations based on the independent measurement of the laser power and the sideband cooling efficiency. Using these fitting parameters, we simulate the cooling for an ion initialized at the Doppler limit and the result is shown in Fig.~\ref{ccool} as a dotted line.  The dashed line is the result of a simulation ignoring the recoil heating in the radial directions.  From this we conclude that this heating does not significantly limit the cooling.  Hence, to get an even lower $\avg{n_z}$, a higher single atom cooperativity is required. This is possible only by improving the experimental setup, such as having a cavity with a better finesse or without birefringence.

\section{Photon statistics} \label{phosta}

\begin{figure}
\includegraphics{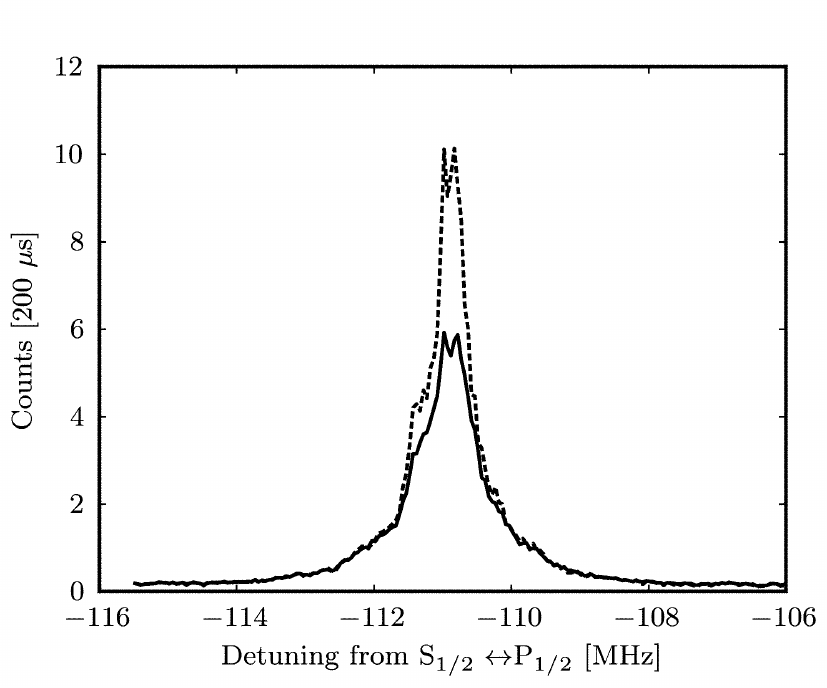}
	\caption{\label{var} A typical ion-cavity emission profile. The solid line is the mean of the photon count $\avg{n}_c$ while the dashed line is the variance $\sigma^2_c$. $\sigma^2_c$ is approximately 1.7 times larger than $\avg{n}_c$ around the cavity peak, indicating super-Poissonian statistics.  }
\end{figure}

In the course of our investigations of cavity cooling it was found that, when the probe laser was tuned near to the cavity resonance, the cavity emission became significantly non-Poissonian, as illustrated in Fig.~\ref{var}. In this section we show that this observation can be explained by accounting for the dependence of the vibrational state on the cavity emission. A theoretical model of phonon-photon coupling under the laser-cavity resonant condition is presented. It is shown that the thermal state of the ions motion can be inferred from the photon distribution and vice-versa and this relationship is demonstrated experimentally over a range of thermal states.

\subsection{Model}

The photon statistics of the spontaneous emission from single atoms/ions driven by cw lasers are known to be sub-Poissonian if the detection window ($\tau$) is small ($\tau \ll 1/\gamma$) and approximately Poissonian if the detection window is large ($\tau \gg 1/\gamma$) \cite{Mandel1,Mandel2}. In the case of ion-cavity coupling, the emitted photons do not follow the same statistical behavior as the coupling is governed by the cavity-induced Raman process. In this process, the photon state is coupled to the vibrational states of the ion which in turn affects the photon statistics. To obtain the phonon-photon statistical relationship, the Hamiltonian of the system is derived. Based on the Hamiltonian, a master equation for the density operator can be established. The master equation can then be solved to obtain a steady state solution for the density operator. Furthermore, the mean photon number and the variance can be computed and the statistical relationship between the phonon states and the photon states can be evaluated.

For a trapped ion beyond LD regime, the spatial extension of the ionic wave function representing the center-of-mass motion (phonon) is no longer small compared with the laser wavelength and results in a coupling between the ion motion and the driving field \cite{vogel,wallen}. To get the Hamiltonian for a trapped ion interacting simultaneously with a intra-cavity field and a laser radiation field, we use the result for a Raman-type interaction from \cite{wallen} which describes the coupling between an ion and two Raman lasers. The expression is modified for the ion-cavity system by replacing a Raman beam with the intra-cavity field. When the cavity is resonant with the laser frequency, the interaction Hamiltonian is
\begin{eqnarray}
\label{hamil}
\hat{H}_{int} &=& \hbar \,\Omega_{R0} \, \hat{f}(\hat{a}_x^{\dag} \hat{a}_x^{};\tilde{\eta}_x) \, \hat{f}(\hat{a}_y^{\dag} \hat{a}_y^{};\tilde{\eta}_y) \, \hat{f}(\hat{a}_z^{\dag} \hat{a}_z^{};\tilde{\eta}_z) \nonumber \\
&\times&  (\hat{c}^{\dag} + \hat{c}) \, ,
\end{eqnarray}
where the operator-valued function $ \hat{f}(\hat{a}_j^{\dag} \hat{a}_j^{};\tilde{\eta_j}) $ is defined in its normally ordered form as
\begin{equation}
\label{foperator}
\hat{f}(\hat{a}_j^{\dag} \hat{a}_j^{};\tilde{\eta}_j) = \exp{[-\tilde{\eta_j}/2]} \, \sum_{l=0}^{\infty} \frac{(i\tilde{\eta_j})^{2l}}{(l!)^2} \, \hat{a}_j^{\dag l} \hat{a}_j^{l} \, .
\end{equation}
In both equations, $\hat{a}_j^{\dag}$ and $\hat{a}_j^{}$ are the phonon creation and annihilation operators, respectively, along the $\hat{\mathbf{j}}$ axis, for $j \in \{x,y,z\}$, while $\hat{c}^{\dag}$ and $\hat{c}$ are the respective cavity photon creation and annihilation operators. The LD parameters, $\tilde{\eta}_j$, are given by
\begin{equation}
\label{neweta}
\tilde{\eta}_x  = \eta_x  \, , \, \tilde{\eta}_y  =  \frac{1}{\sqrt{2}} \, \eta_y  \, , \, \tilde{\eta}_z  =  \frac{1}{\sqrt{2}} \,\eta_z \, ,
\end{equation}
where $\eta_j = k \sqrt{\hbar/(2m\omega_j)}$, $k$ is the wavenumber and $m$ is the mass of the ion. The von Neumann equation for the density operator ($\hat{\rho}$) is
\begin{equation}
\label{rho}
\frac{\textrm{d} \hat{\rho}}{\textrm{dt}} = -\frac{i}{\hbar} \, [\hat{H}_{int} \, ,\hat{\rho}] +  \mathcal{L}(\hat{\rho}) \, ,
\end{equation}
where $\mathcal{L}$ is the superoperator,
\begin{equation}
\label{super}
\mathcal{L}(\hat{\rho}) \, = \kappa ( 2 \, \hat{c}^{\dag} \hat{ \rho} \hat{c} - \hat{ \rho}  \hat{c}^{\dag} \hat{c} -\hat{c}^{\dag} \hat{c} \hat{ \rho}  ) \, .
\end{equation}
Since $\ket{n}$ is an eigenstate of $\hat{f}(\hat{a}^{\dag} \hat{a}^{};\tilde{\eta})$ with eigenvalue $\exp{[-\tilde{\eta}^2/2]} \, L_n (\tilde{\eta}^2)$, the steady state solutions of Eq.~(\ref{rho}) are of the form
\begin{eqnarray}
\label{sssol}
&&\ket{n_x } \bra{n_x} \otimes \ket{n_y} \bra{n_y}\otimes \ket{n_z} \bra{n_z} \nonumber \\
&&\otimes \ket{\alpha_{n_x,n_y,n_z}} \bra{\alpha_{n_x,n_y,n_z}} \, ,
\end{eqnarray}
where $\ket{n_j }$ are phonon Fock state along $\hat{\mathbf{j}}$ axis and $\ket{\alpha_{n_x,n_y,n_z}}$ are photon coherent states with amplitude given by
\begin{eqnarray}
\label{alpha}
\alpha_{lmn} &=&  \frac{\Omega_{R0}}{i \kappa}  \exp{[-(\,\tilde{\eta}_x^2+\tilde{\eta}_y^2+\tilde{\eta}_z^2\,)/2]} \nonumber \\
& & \times L_l(\tilde{\eta}_x^2) \, L_m(\tilde{\eta}_y^2) \, L_n(\tilde{\eta}_z^2).
\end{eqnarray}
We assume a thermal distribution of phonons and so the reduced steady state density operator for the intra-cavity field is given by
\begin{eqnarray}
\label{density}
\hat{\rho}_{ss} &=& \sum_{l,m,n}  \, \rho_{lmn}  \, \ket{\alpha_{l,m,n}}\bra{\alpha_{l,m,n}} \,  ,
\end{eqnarray}
with
\begin{equation}
\label{rho2}
\rho_{lmn} = \frac{\avg{n_x}^l}{(1+ \avg{n_x})^{l+1}} \frac{\avg{n_y}^m}{(1+ \avg{n_y})^{m+1}} \frac{\avg{n_z}^n}{(1+ \avg{n_z})^{n+1}} \, .
\end{equation}
The average intra-cavity photon number, $\avg{n}_c$, is then given by
\begin{eqnarray}
\label{photonmean}
\avg{n}_c &=& \textrm{Tr} \{\hat{\rho}_{ss} \hat{c}^{\dag}\hat{c} \}  \nonumber \\
 &=& \,\sum_{l,m,n} \, \rho_{lmn} |\alpha_{lmn}|^2\,
\end{eqnarray}
and similarly
\begin{eqnarray}
\label{photonsecond}
\avg{n^2}_c &=& \textrm{Tr} \{\hat{\rho}_{ss} (\hat{c}^{\dag}\hat{c})^2 \}  \nonumber \\
 &=& \sum_{l,m,n}  \, \rho_{lmn}  \Big( |\alpha_{lmn}|^2+|\alpha_{lmn}|^4 \Big)\,  .
\end{eqnarray}
With Eq.~(\ref{photonmean}) and Eq.~(\ref{photonsecond}), it is straight forward to compute the variance $\sigma_c^2=\avg{n^2}_c -\avg{n}_c^2$.

The photon statistics can be characterized by the Fano factor $F$ \cite{fano} which is the ratio of variance $\sigma^2$ to the mean $\avg{n}$.  In our case this is given by
\begin{eqnarray}
\label{fanofac}
F &=& \frac{\sigma_c^2}{\avg{n}_c} \nonumber \\
&=& 1 + \frac{ \sum \rho_{lmn} |\alpha_{lmn}|^4}{ \sum \rho_{lmn} |\alpha_{lmn}|^2}  - \sum \rho_{lmn} |\alpha_{lmn}|^2 \, ,
\end{eqnarray}
where all summations are over the indices $l$, $m$ and $n$.

The photon distribution for each individual $n$-vibrational state is Poissonian as given by the properties of coherent states. Thus, the effective photon distribution is given by the sum of Poissonians weighted according to the thermal distribution. When the ion is cold and the phonon population is distributed only around the motional ground state, the photon distribution is near Poissonian. As $\avg{n}$ increases, $F$ increases indicating a super-Poissonian distribution.

\subsection{Experiments}

To obtain the Fano factor as a function of $\avg{n_z}$, we performed experiments on a single $^{138}\textrm{Ba}^+$ ion prepared at several values of $\avg{n_z}$. The control of $\avg{n_z}$ was achieved by varying the parameters of the Doppler cooling lasers. For $\avg{n_z}$ smaller than the Doppler limit ($\avg{n_z} \approx 25$), Raman sideband cooling was used.

In each experimental cycle, a Doppler cooling pulse was turned on for $1 \, \mathrm{ms}$. This pulse was followed by $3 \, \mathrm{ms}$ of Raman sideband cooling for the measurement with $\avg{n_z}$ smaller than the Doppler limit. The ion-cavity emission was then observed while probing with beam $R_p$. To avoid heating of the ion during probing, this beam was switched on for only $100 \,\mu\textrm{s}$. During each probing event, the emitted photons were counted with a single photon counting module (SPCM) coupled to the output of the cavity via a single mode fiber.  From the transmission of the cavity at $493\,\mathrm{nm}$ ($24\%$), the fibre coupling efficiency ($70\%$), and the quantum efficiency of the SPCM at $493\,\mathrm{nm}$ ($45\%$) we estimate an overall detection efficiency of intra-cavity photons of approximately $7.5\%$.

Each statistical photon distribution was obtained from 1000 probing events, yielding a single value for the Fano factor. $300$ sets of these distributions were collected in order to acquire sufficient samples of mean and variance values to compute the dispersion of the Fano factors for each $\avg{n_z}$.

\subsection{Results}

\begin{figure}
\includegraphics{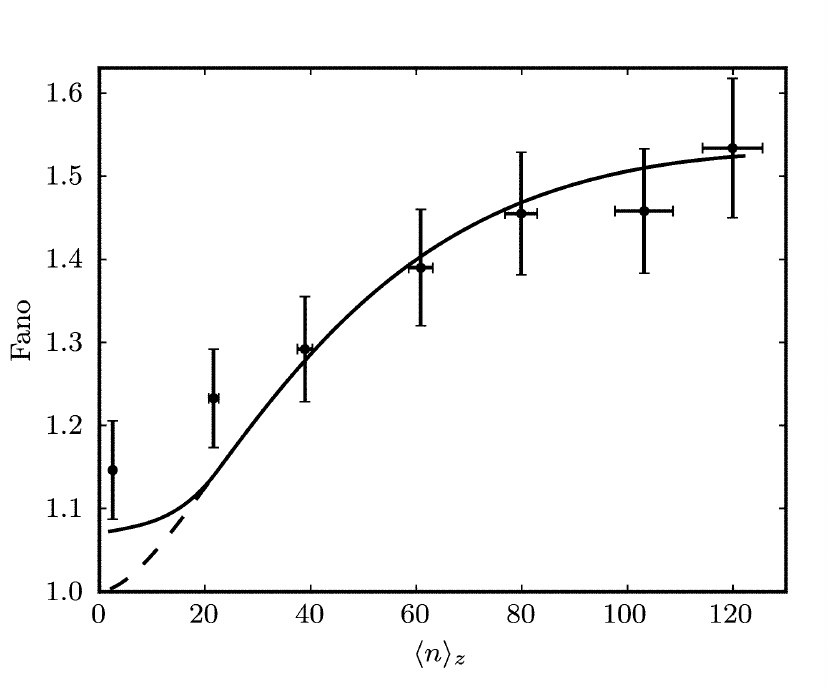}
	\caption{\label{fanoplot} Fano factor $F$ as a function of the averaged phonon number $\avg{n_z}$. The solid line is the fit of the theoretical model to the experimental data while the dashed line simulates the ideal situation where the temperatures of the ion are equivalent in all dimensions. The $F$ error bars are the standard deviation of the Fano factor over 300 sets of photon statistics while the $\avg{n_z}$ error bars are the uncertainties from the temperature measurements. Raman sideband cooling was used for the first data point such that the $\avg{n_z}$ is brought down to $\approx 2$. Note that the sideband cooling is only along $z$ axis, hence the ion temperature in other axes should be higher than the Doppler limit. This is a one-parameter fit with the fitted value of $\Omega_{R0}= 2 \pi \times 0.14(1) \, \textrm{MHZ}$. The reduced $\chi^2$ of the fit is $\approx$ 0.8.}
\end{figure}

Fitting the experimental data to the theoretical model requires the information on phonon occupation numbers of all vibrational modes. Due to the current configuration (see Sec.~\ref{cavexp}), sideband cooling and temperature measurements can be performed only along the $\hat{\mathbf{z}}$ axis. However, the ion is assumed to be in thermal equilibrium after Doppler cooling so that the temperature along the other axes is also at the Doppler limit. We therefore use the relationship
\begin{equation}
\label{trapfreq}
\avg{n_x} \approx \avg{n_y} \approx  \frac{\omega_z}{\omega_r} \avg{n_z} \, ,
\end{equation}
which reduces the independent variables to only $\avg{n_z}$ and $\Omega_{R0}$. In the measurement where sideband cooling is used, the temperature along the non-cavity cooled axes is higher than the Doppler limit. However, the additional heating can be properly accounted using the approach in Sec.~\ref{cavexp}.

Both the experimental data and the fit are shown in Fig.~\ref{fanoplot}. Eq.~(\ref{fanofac}) is fitted to the experimental data where the only free parameter is the ion-cavity coupling rate $\Omega_{R0}$. The fitted value is $2 \pi \times 0.14(1) \, \textrm{MHz}$, which corresponds to a laser Rabi rate of $\Omega_{L0}= 2 \pi \times 20(1) \, \textrm{MHz}$ and agrees with an estimation based on the measured power of the $R_p$ laser. Minimizing the $\chi^2$ of the fit produces a reduced $\chi^2$ of 0.8, indicating that the experimental outcome is in good agreement with the model prediction.

The experimental results confirm that the Fano factor is related to the ion temperature as described by the model. However, as seen in Fig.~\ref{fanoplot}, the  span of the error bars reduces the precision in deducing the ion temperature for a given Fano number. The Fano error bars in the figure are mainly due to the intensity fluctuation of $\pm 5\%$ at the cavity output. This is caused by the frequency drift in the Doppler cooling laser and the drift of the ion position from the cavity anti-node  throughout the experimental period, which slightly affect the ion temperature as well as the ion-cavity coupling efficiency. Although the error bars are only a few percent compared to the signal amplitude, the error bars between the adjacent data points are still overlapping due to the small gradient curve in the figure. A better precision is achievable by improving the laser reference stability which results in a smaller error bar or having a larger ion-cavity coupling rate, $\Omega_{R0}$, which results in a steeper curve. As increasing the laser intensity $\Omega_{L0}$ would induce undesired ion heating, a larger $\Omega_{R0}$ can only be achieved by improving the ion-cavity coupling strength ($g_0$). However, this is not readily achievable with our current setup.

The experiment demonstrates the possibility of using the photon statistical distribution as a temperature probe. The advantage of implementing this technique is that the ion temperature can potentially be measured within a relatively short period of time. Applying the typical measurement methods as discussed in Sec.~\ref{tempmeasure} or in \cite{walther} usually require long experimental times to acquire sufficient data. However, the method proposed here requires only one statistical distribution to estimate the Fano factor for a particular photon distribution. Thus, the ion temperature can be estimated in almost a real-time manner: less than a second in our experiment to acquire a statistical distribution. This is particularly useful for setup and optimization of processes such as Doppler cooling in which rapid monitoring of the ion temperature is definitely helpful.

\section{Conclusions}

We have demonstrated cavity cooling of an ion to sub-Doppler temperatures for the first time. A theoretical model which describes the cooling dynamics and the limiting factors is presented and validated by the experimental results. Useful information, such as the effect of recoil heating on the cooling performance, are provided and may help future efforts in achieving ground state cavity cooling in an ion-trap.

In the second part of the paper, we have shown the connection between the statistical distribution of the cavity photon emission and the ion temperature. This potentially allows for the use of the photon statistics as a fast temperature probe for a trapped ion.


\begin{acknowledgments}
We thank Markus Baden, Kyle Arnold and Andrew Bah for help with preparing the manuscript. This research was supported by the National Research Foundation and the Ministry of Education of Singapore.
\end{acknowledgments}

\appendix

\section{Thermal effect on ion-cavity coupling} \label{classicLD}

To estimate the thermal effect on ion-cavity coupling, we consider the setup depicted in Fig.~\ref{iso} in which a single ion is located at $\vec{\textbf{r}}= x \,\hat{\textbf{x}}+ y \,\hat{\textbf{y}} +z \,\hat{\textbf{z}}$. The ion interacts with a far detuned laser radiation field $R_p$ and an orthogonal cavity field. The ion-cavity Raman rate $\Omega_R$ due to the cavity induced Raman transition is
\begin{eqnarray}
\label{cavrate}
\frac{\Omega_L(\vec{\textbf{r}}) \, g(\vec{\textbf{r}})}{\Delta} &=& \frac{ \Omega_{L0} \, g_0}{\Delta} \sin{(k x + \phi_c)}  \nonumber \\
&\times&\exp{\bigg[ik\frac{y+z}{\sqrt{2}} \bigg]} \exp{\bigg[-\frac{y^2+z^2}{w_L^2}\bigg]} \, ,
\end{eqnarray}
where $\Omega_L(\vec{\textbf{r}})$ is the laser Rabi rate with a maximum value of $\Omega_{L0}$ at the center of the mode waist, $ g(\vec{\textbf{r}})$ is the ion-cavity coupling rate with a maximum value of $g_0$ at the cavity anti-node, $\Delta$ is the detuning of the laser frequency from the ionic resonance, $k$ is the wavenumber, $\phi_c$ is the phase of the intra-cavity standing wave and $w_L$ is the beam waist of the laser. Due to the position spread of the ion in the trap, this Raman rate has to be averaged over the Gaussian localization of the ion wavepacket
\begin{equation}
\label{omer}
\Omega_R= \int{\rho(\vec{\textbf{r}})}\, \frac{\Omega_L(\vec{\textbf{r}}) \, g(\vec{\textbf{r}})}{\Delta}\textrm{ d}\vec{\textbf{r}} \, ,
\end{equation}
where
\begin{eqnarray}
\label{wavepac}
\rho(\vec{\textbf{r}}) &=& \frac{1}{(2 \pi)^{3/2} \, \sigma_x\sigma_y\sigma_z}  \nonumber \\
&\times&\exp{\bigg[-\frac{x^2 }{2 \sigma_x^2}\bigg]} \exp{\bigg[-\frac{y^2 }{2 \sigma_y^2}\bigg]}\exp{\bigg[-\frac{z^2 }{2 \sigma_z^2}\bigg]}\, .
\end{eqnarray}
Here $\sigma_j$ is defined as the wave function spread of the ion along the $\hat{\mathbf{j}}$ axis
\begin{equation}
\label{sigma3}
\sigma_j = \sqrt{\frac{k_B T}{m \, \omega_j^2}} \, .
\end{equation}
Approximating $\omega_x \approx \omega_y \approx \omega_r$, $\Omega_R$ is then evaluated as
\begin{eqnarray}
\label{omer2}
\Omega_R &=& \Omega_{R0} \, \frac{\sigma_r'\sigma_z'}{\sigma_r\sigma_z} \, \sin{(\phi_c)} \nonumber \\
&\times&  \exp{\bigg[-\frac{(k \sigma_r')^2}{4}-\frac{(k \sigma_z')^2}{4}-\frac{(k \sigma_r)^2}{2}\bigg]} \, ,
\end{eqnarray}
with
\begin{equation}
\Omega_{R0} = \frac{\Omega_{L0} \, g_0}{\Delta}
\end{equation}
and
\begin{equation}
\sigma_j'^2 = \frac{(\sigma_j w_L)^2}{w_L^2+2\,\sigma_j^2} \, .
\end{equation}
If we consider that the ion is positioned at the cavity anti-node and the laser beam waist is much larger than the wave function spread, $\phi_c = \pi/2$ and $\sigma_j' \approx \sigma_j$. Eq.~(\ref{omer2}) can thus be simplified to
\begin{equation}
\label{omer3}
\Omega_R= \Omega_{R0}  \exp{\bigg[-\frac{3(k \sigma_r)^2}{4}-\frac{(k \sigma_z)^2}{4} \bigg]} \, .
\end{equation}
To obtain a Raman rate with explicit dependency on temperature, $\sigma_j$ can be written in terms of the LD parameter $\eta_j$ and the average phonon occupation number $\avg{n_j}$
\begin{eqnarray}
\label{sigma}
k \sigma_j &=& k \sqrt{\frac{\hbar (\avg{n_j}+1/2) }{m \omega_j}} \\
\label{sigma2}
 &=& \eta_j \sqrt{2(\avg{n_j}+1/2)}\, ,
\end{eqnarray}
with $\eta_j = k \sqrt{\hbar/(2m\omega_j)}$. Substituting Eq.~(\ref{sigma2}) into Eq.~(\ref{omer3}), a Raman rate with thermal effect incorporated is obtained
\begin{eqnarray}
\label{omer4}
\Omega_R &=& \Omega_{R0}  \exp{\bigg[-\frac{3\eta_r^2 (\avg{n_r}+1/2)}{2} \, \bigg]} \nonumber \\
&\times& \exp{\bigg[-\frac{\eta_z^2 (\avg{n_z}+1/2)}{2} \, \bigg]} \, .
\end{eqnarray}
The single atom cooperativity determines the scattering rate into the cavity. Since this scattering rate is proportional to $\Omega_R^2$, we use the following expression for the thermally averaged cooperativity
\begin{eqnarray}
\label{effcoop}
\widetilde{C}\,(\avg{n_r},\avg{n_z}) & =&  C \exp{\big[-3\eta_r^2 (\avg{n_r}+1/2) \, \big]} \nonumber \\
&\times& \exp{\big[-\eta_z^2 (\avg{n_z}+1/2) \, \big]} \, ,
\end{eqnarray}
which converges to the single atom cooperativity $C$ for an ion in the LD regime $\eta_j^2 \avg{n_j} \ll 1$. For a trapped ion beyond the LD regime, we use Eq.~(\ref{effcoop}) for the effective cooperativity at a given temperature.

\section{Recoil heating vs cavity cooling} \label{recoilheating}

To model our results, we first note that the motional sidebands are well resolved by the cavity, thus cooling can be selectively performed along a single direction. Other directions are then only affected by recoil heating. For cooling along a particular direction, namely the $\hat{\mathbf{z}}$ axis, the rate equation for the $n$-phonon occupation probability $P_{n}$ is \cite{morigi,stenholm}
\begin{eqnarray}
\label{prate}
\frac{\textrm{d}}{\textrm{dt}} P_{n_z} =  \eta_z^2   &\Big\{ &\left(n_z+1\right) A_{-} P_{n_z+1} \nonumber \\
&-& \Big[ \left(n_z+1\right) A_{+}+ n_z A_{-} \Big] P_{n_z}  \nonumber \\
&+& n_z A_{+} P_{n_z-1} \Big\}\, ,
\end{eqnarray}
where $A_{-}$ and $A_{+}$ are the cooling and the heating rate, respectively. We use the rate equations for the good-cavity regime in \cite{morigi} where the laser detuning $\Delta$ is much larger than the total dipole decay rate $\gamma=\Gamma/2$, the ion-cavity coupling strength $g$ and the cavity field decay rate $\kappa$.

In \cite{morigi}, the rates are derived presuming a stationary state solution. However, when the cooling is only along the axial direction, the radial phonon occupation number $\avg{n_r}$ increases over time due to recoil heating,
\begin{equation}
\label{heatnr}
\frac{\textrm{d}}{\textrm{dt}}\avg{n_r} \approx \gamma \frac{\Omega_{L0}^2}{2 \Delta^2} \eta_r^2 \, .
\end{equation}
Consequently, as seen in Eq.~(\ref{effcoop}), the cooperativity decreases over time and results in a smaller photon scattering rate into the cavity. This eventually leads to a time dependence of the cooling and heating rates. Thus, $A_{\pm}$ rates never stabilize to stationary values due to their dependence on $\avg{n_r}$ and no steady-state temperature is achieved.

Although the stationary state condition is not fulfilled here, we assume a pseudo steady-state which depends on the time-varying thermally averaged cooperativity. Thus, the resultant rates vary according to the thermally averaged cooperativity,
\begin{equation}
\label{coolandheat1}
A_{\pm } =  \gamma \, \frac{  \Omega_{L0}^2}{2 \Delta^2} \left[ \alpha+\varphi_l^2 +\frac{\widetilde{C}}{2} \, \frac{\kappa^2}{\kappa^2+ \left(\delta_c\mp\omega_z \right)^2} \right] \, ,
\end{equation}
where $\alpha$ is a geometric factor (1/3 for $\mathrm{J}'_{1/2} \rightarrow \mathrm{J}_{1/2}$ transition), $\varphi_l$ is the cosine of the angle between the driving laser and $\hat{\mathbf{z}}$ axis and $\delta_c$ is the relative detuning between the laser and the cavity resonance.

In addition, we need to account for the cavity birefringence. Because of our limited optical access, we can only probe at an angle of $45^\circ$ to the vertical ($\hat{\mathbf{y}}$). Thus the probe couples equally to both birefringence modes of the cavity. In our case the cavity birefringence is not well resolved, and the best cooling rate is obtained when the cooling laser is tuned such that $\delta_c = -\omega_z \pm \delta_b$, where $2 \delta_b$ is the separation between the modes. In this case, the heating rate is not strongly affected by the birefringence and is approximately the same as that in Eq.~(\ref{coolandheat1}), namely
\begin{equation}
\label{coolandheat2}
A_{+} \approx  \gamma \, \frac{  \Omega_{L0}^2}{2 \Delta^2} \left[ \alpha+\varphi_l^2 +\frac{\widetilde{C}}{2} \, \frac{\kappa^2}{\kappa^2+ \left(2 \omega_z \right)^2} \right] \, .
\end{equation}
As for the cooling rate, the substitution of $\delta_c$ into Eq.~(\ref{coolandheat1}) leads to
\begin{equation}
\label{coolandheat3}
A_{-} =  \gamma \, \frac{  \Omega_{L0}^2}{2 \Delta^2} \left[ \alpha+\varphi_l^2 +\frac{\widetilde{C}}{2} \, \frac{\kappa^2}{\kappa^2+ \delta_b ^2} \right] \, ,
\end{equation}
which translates into a lower cooling rate compared to a non-birefringent cavity.

Multiplying Eq.~(\ref{prate}) by $n_z$ and summing over all $n_z \geq 0$ results in a rate equation for $\avg{n_z}$
\begin{equation}
\label{nrate}
\frac{\textrm{d}}{\textrm{dt}} \avg{n_z} =  -\eta_z^2  \Big[ A_{-} \avg{n_z} - A_{+} (\avg{n_z}+1) \Big]\, .
\end{equation}

An analytical solution for $\avg{n_z}$ can be obtained by assuming a stationary state condition where $A_{+} P_{n_z} = A_{-} P_{n_z+1}$ for $t \rightarrow \infty$ \cite{morigi,stenholm}. Unfortunately, this condition is not achievable with our current setup. Nevertheless, even without an analytical solution, both Eq.~(\ref{heatnr}) and Eq.~(\ref{nrate}) are solved jointly and numerically and used for fitting to the experimental data.

\bibliography{ref2}

\end{document}